\documentclass[preprint,showpacs,preprintnumbers,amsmath,amssymb]{revtex4}
\usepackage{booktabs}
\usepackage{mathrsfs}
\usepackage{epsfig}
\usepackage{graphicx}
\usepackage{dcolumn}
\usepackage{bm}
\usepackage{amsmath}
\usepackage{slashed}       
\usepackage{amssymb}
\usepackage{amsfonts}
\usepackage{graphics,float,psfrag}   
\usepackage[usenames,dvips]{color}
\usepackage{rotating}

\let\jnfont=\rm
\def\NPB#1,{{\jnfont Nucl.\ Phys.\ B }{\bf #1},}
\def\PLB#1,{{\jnfont Phys.\ Lett.\ B }{\bf #1},}
\def\EPJC#1,{{\jnfont Eur.\ Phys.\ Jour.\ C }{\bf #1},}
\def\PRD#1,{{\jnfont Phys.\ Rev.\ D }{\bf #1},}
\def\PRL#1,{{\jnfont Phys.\ Rev.\ Lett.\ }{\bf #1},}
\def\MPLA#1,{{\jnfont Mod.\ Phys.\ Lett.\ A }{\bf #1},}
\def\JPG#1,{{\jnfont J.\ Phys.\ G}{\bf #1},}
\def\CTP#1,{{\jnfont Commun.\ Theor.\ Phys.\ }{\bf #1},}
\def\ZPC#1,{{\jnfont Z.\ Phys.\ C }{\bf #1},}
\def\JHEP#1,{{\jnfont JHEP \ }{\bf #1},}
\def\lsim{\raise0.3ex\hbox{$<$\kern-0.75em\raise-1.1ex\hbox{$\sim$}}}
\def\gsim{\raise0.3ex\hbox{$>$\kern-0.75em\raise-1.1ex\hbox{$\sim$}}}

\begin{document}
\preprint{\parbox{1.2in}{\noindent arXiv:}}

\title{The Heavy Higgs Decay Effect and Its Impact on Higgs Data Fitting in NMSSM}

\author{Mengchao Zhang}

\affiliation{
   State Key Laboratory of Theoretical Physics,
      Institute of Theoretical Physics, Academia Sinica, Beijing 100190, China
      \vspace{1cm}}

\begin{abstract}
Besides the loop contribution and the coupling changing, heavy Higgs decaying to 125GeV Higgs is another important effect in Higgs property measurement in the models that have extend Higgs sector.
Such heavy Higgs decay effect contain more than one sources, such as $A\rightarrow Zh$, $H\rightarrow hh$, $A_2\rightarrow A_1h$, and $H_3\rightarrow H_1H_2$.
Some of them will be promising in NMSSM Higgs sector searching.
Higgs data fitting is performed with and without this heavy Higgs decay effect being included, and the final result demonstrate such effect is very important in Higgs data fitting if the signal efficiency is large.
\end{abstract}
\maketitle

\section{Introduction}
With a 125GeV Higgs-like scalar found by the LHC\cite{ATLAS-Higgs}\cite{CMS-Higgs}, the last cornerstone of the Standard Model(SM) have been confirmed.
The measurement of the scalar's couplings to vector boson and fermions are compatible with the SM predictions\cite{HiggsUpdatedMeasurementATLAS}\cite{HiggsUpdatedMeasurementCMS}.
Even so, due to some difficulty like the gauge hierarchy, dark matter, or vacuum stability, it is commonly believed that SM is just an effective theory in low energy scale and New Physic(NP) will emerge in some higher energy scale.
While the LHC haven't find any unambiguous NP signals, so a main work at present is using the low energy observables, especially the Higgs data obtained these years, to constrain the NP models' parameter space.
A convenient and popular method to perform the Higgs data constraints is through the Higgs signal strength fitting.
Higgs signal strength is the ratio of observed Higgs signal number to SM predicted Higgs signal number, and Higgs signal number is the product of Higgs production crosssection, branching ratio, and signal efficiency.
If the NP model have a light particle that can change the Higgs effective coupling to photon and gluon in the loop level, or have a scalar that can mix with the 125GeV SM-like Higgs, or lead to a non-SM Higgs decay channel, then the Higgs signal strength will change, then one can use the observed Higgs signal strength to exclude the NP points that leading to a too large or too small signal strength.
In such situation, the topology of the Higgs process are just the same as the SM Higgs process.
It means, you don't need to worry about the signal efficiency, because signal efficiency have been cancelled completely in the ratio(the signal efficiency in the numerator and denominator are equalled).
Such argument is true when the non-SM Higgs production crosssection is negligible compared to the SM Higgs production crosssection.
But, if the non-SM Higgs production crosssection and SM Higgs production crosssection are on the same level, then the effective coupling and branching ratio are not enough for the Higgs signal strength fitting. The signal mixing of the SM produced Higgs and non-SM produced Higgs should be included.
While such a possibility have been analysed in Minimal Supersymmetry Standard Model(MSSM)\cite{Yu:2014mda}(Higgs and neutralino associated production) and two Higgs doublet model(2HDM)\cite{Higgs_Chain}(SM-like Higgs are decayed from heavier scalars).
In this paper we will analyse the non-SM Higgs production in the Next to Minimal Supersymmetry Standard Model(NMSSM) and its influence on Higgs data fitting.
Now let me explain why we choose NMSSM.

Among various New Physic models, Supersymmetry(SUSY) is a very attractive one due to its elegant solution to hierarchy problem and other theoretical and phenomenological merit like providing dark matter candidate, unifying the gauge coupling, or stabilizing the vacuum in high energy scale.
The most studied and popular low energy scale SUSY model is MSSM.
In MSSM, there are 5 Higgs states(2 CP-even neutral scalars, 1 CP-odd neutral scalar and a pair of charged scalar) after spontaneous symmetry breaking, and the lighter CP-even scalar can be regarded as the 125GeV scalar found in LHC.
But the up limit of the lighter CP-even scalar mass in MSSM is about 130GeV\cite{uplimit}, so the 125GeV Higgs will leads to some degree of fine-tuning\cite{125finetunning}.
The $\mu$ term in MSSM super potential also cause some theoretical puzzle.
While in NMSSM, fine-tuning can be alleviated by an additional singlet which can enhance the tree level Higgs mass\cite{King_NMSSM}.
At the same time, the puzzling $\mu$ term can be explained by a dynamical origin.
More paper related to naturalness, see .
On the other hand, if the non-SM Higgs production process is heavier scalar's decay, then there always need a small $\tan\beta$\cite{Higgs_Chain}\cite{Djouadi_low_tanb}.
It is partly because small $\tan\beta$ will lead to a small branching ratio to bottom quark pair.
Besides, if the heavy scalar is heavier than 2 times of top quark mass, then the branching ratio to top pair will be dominant and the branching ratio to 125GeV Higgs will be strongly supressed.
Such an small $\tan\beta$ and light non-SM Higgs scenario is unfavoured in MSSM but can be easily obtained in NMSSM\cite{Wu:2015nba}\cite{TaoHanNMSSM}(because the Higgs mass enhancement in tree-level and the mixing with the singlet).
So the NMSSM Higgs sector is very hopeful in non-SM Higgs production.

In this paper, we will analyse the production and decay of the NMSSM Higgs sector in detail, show how large this non-SM Higgs production crosssection can be, and estimate the impact of this non-SM Higgs production on Higgs data fitting.
In next section, we will briefly review the NMSSM Higgs sector.
In section 3, we will detailedly describe our scan method.
Section 4 is used to show the results, and we analyse the Higgs data fitting including non-SM Higgs production in section 5.
Section 6 is the conclusion.

\section{NMSSM Higgs sector}
The NMSSM Higgs sector contain two Higgs doublets and a Higgs singlet\cite{Ellwanger:2009dp}:
\begin{equation}\
  H_u = \left(
          \begin{array}{c}
            H_u^+ \\
            H_u^0 \\
          \end{array}
        \right),
          H_d = \left(
          \begin{array}{c}
            H_d^0 \\
            H_d^- \\
          \end{array}
        \right),
        S.
\end{equation}
Without regard to domain wall problem, the $\mathbb{Z}$3-invariant NMSSM superpotential and corresponding Higgs soft breaking terms are\cite{Ellis_potential}\cite{Miller_HiggsSector}:
\begin{eqnarray}
  W &=& \hat{u}^c\mathbf{h_u}\hat{Q}\hat{H}_u - \hat{d}^c\mathbf{h_d}\hat{Q}\hat{H}_d - \hat{e}^c\mathbf{h_e}\hat{L}\hat{H}_d
  + \lambda\hat{S}\hat{H}_u\hat{H}_d + \frac{1}{3}\kappa\hat{S}^3,\\
  -\mathcal{L}_{soft} &=& m_{H_u}^2|H_u|^2 + m_{H_d}^2|H_d|^2 + m_{S}^2|S|^2 + [\lambda A_{\lambda}SH_{u}H{d} +
  \frac{1}{3}\kappa A_{\kappa}S^3 + h.c.].
\end{eqnarray}
After spontaneous symmetry breaking(SSB), Higgs fields will get vacuum expected values:
\begin{equation}\
  \langle H_u\rangle = \frac{1}{\sqrt{2}}\left(
          \begin{array}{c}
            0 \\
            v_u \\
          \end{array}
        \right),
          \langle H_d\rangle = \frac{1}{\sqrt{2}}\left(
          \begin{array}{c}
            v_d \\
            0 \\
          \end{array}
        \right),
        \langle S\rangle = \frac{1}{\sqrt{2}}v_s,
\end{equation}
with $v\equiv\sqrt{v_u^2 + v_d^2}=246GeV$.

Using three minimum conditions, $m_{H_u}^2$, $m_{H_d}^2$ and $m_{S}^2$ can be represented by other parameters, then the independent parameters used to describe NMSSM Higgs sector are: $\lambda$, $\kappa$, $\mu = \lambda v_s$, $\tan\beta = v_u/v_d$, $A_{\lambda}$, $A_{\kappa}$. Rotating the unphysical goldstone state away, the mass matrix of pseudoscalar is:
\begin{eqnarray}
  M_{P11}^2 &=& M_A^2, (M_A^2 \equiv \frac{\lambda v_s}{\sin2\beta}(\sqrt{2}A_{\lambda} + \kappa v_s)) \\
  M_{P12}^2 &=& \frac{1}{2}(M_A^2\sin2\beta-3\lambda\kappa v_s^2)\frac{v}{v_s}    ,\\
  M_{P22}^2 &=& \frac{1}{4}(M_A^2\sin2\beta+3\lambda\kappa v_s^2)\frac{v^2 \sin2\beta}{v_s^2}-3\kappa v_s A_{\kappa}/\sqrt{2}.
\end{eqnarray}
Our work is concentrate on the Higgs sector, so it will be more convenient to replace $A_{\lambda}$ and $A_{\kappa}$ by $M_A\equiv M_{P11}$ and $M_P\equiv M_{P22}$ in NMSSM Higgs sector description.
$M_A$ and $M_P$ are the mass scale of heavier doublet and singlet separately.
But unlike MSSM, $M_A$ and $M_P$ are just two approximate mass scale in NMSSM, not physical observables.

Then the scalar mass matrix is:
\begin{eqnarray}
  M_{S11}^2 &=& M_A^2+(M_Z^2-\frac{1}{2}{\lambda}^2{v}^2){\sin}^22\beta ,\\
  M_{S12}^2 &=& -\frac{1}{2}(M_Z^2-\frac{1}{2}{\lambda}^2{v}^2)\sin4\beta ,\\
  M_{S13}^2 &=& -\frac{1}{2}(M_A^2\sin2\beta+\lambda\kappa v_s^2)\frac{v\cos2\beta}{v_s} ,\\
  M_{S22}^2 &=& M_Z^2\cos^2 2 \beta +\frac{1}{2} (\lambda v)^2 \sin^2 2 \beta ,\\
  M_{S23}^2 &=& \frac{1}{2}(2\lambda^2v_s^2-M_A^2\sin^22\beta-\lambda\kappa v_s^2\sin2\beta)\frac{v}{v_s} ,\\
  M_{S33}^2 &=& \frac{1}{4}M_A^2\sin^22\beta \frac{v^2}{v_s^2}+2\kappa^2v_s^2+\kappa v_s A_{\kappa}/\sqrt{2}-\frac{1}{4}\lambda\kappa v^2\sin2\beta.
\end{eqnarray}
The charged Higgs mass is:
\begin{eqnarray}
  M_{H\pm}^2 &=& M_A^2+M_W^2-\frac{1}{2}(\lambda v)^2.
\end{eqnarray}
Where $M_Z$ and $M_W$ are the mass of Z boson and W boson separately.
The scalar mass eigenstates are denoted by $H_1$, $H_2$, and $H_3$ with increasing mass.
The pseudoscalar mass eigenstates are denoted by $A_1$ and $A_2$ with $A_2$ is heavier than $A_1$.
Charged Higgs is denoted by $CH$.

A analytical study can be performed in MSSM Higgs sector(tree-level) if $M_A$ is heavy or the $\tan\beta$ is large than 5, because the Higgs spectrum and the Higgs mixing angle can be expressed by $\frac{v}{M_A}$ or $\frac{1}{\tan\beta}$.
While in NMSSM, $M_A$ and $M_P$ maybe not too larger or smaller than $v$, and $\tan\beta$ tend to be small, so such an expansion is impossible in most situation.
In addition, NMSSM Higgs sector is described by 6 parameters, and the physical observables' dependence on these 6 parameters are complicated.
So we will scan the parameter space that leading to what interests us, and use a lot of result to show the  property of this parameter space.
In next section we will describe the scan method in detail.

\section{Scan Method}
We use $\verb"NMSSMTools4.5.1"$\cite{NMSSMTools} to scan the parameter space.
Package $\verb"NMHDECAY"$\cite{NMHDECAY} and $\verb"NMSDECAY"$\cite{NMSDECAY}, based on $\verb"HDECAY"$\cite{HDECAY} and $\verb"SDECAY"$\cite{SDECAY}, are used to calculate the decay widths and branching ratios of Higgs and sparticle.
The NMSSM particle mass spectrum(include Higgs mass), mixing angle, and couplings are also calculated by $\verb"NMSSMTools4.5.1"$.

\subsection{Scan range}
The value range of the 6 parameters mentioned before can be confined by a series of conditions such like vacuum stability or perturbativity in high energy scale\cite{Miller_HiggsSector}. Combined with the argument in the introduction, we set our parameters range as below:
\begin{itemize}

  \item In order to avoid Landau Pole before the Great Unification scale, $\lambda$ and $\kappa$ can't be too large:
  \begin{equation}
      0.0\leq\lambda,\kappa\leq0.7.
  \end{equation}

  \item As we mentioned in the introduction, small $\tan\beta$ is favoured by large Higgs decay branching ratio, so:
  \begin{equation}
      1.0\leq\tan\beta\leq5.0.
  \end{equation}

  \item In MSSM, the lower limit of $M_A$ is about 300GeV(except in the alignment region\cite{Carena:2014nza}). While in NMSSM, due to the mixing with SM singlet, the lower limit of $M_A$ can reach 200GeV. Besides, we don't want too heavy non-SM Higgs, so:
  \begin{equation}
      200GeV\leq M_A \leq500GeV.
  \end{equation}

  \item While the singlet mass $M_P$ can vary in a larger range:
  \begin{equation}
      0GeV\leq M_P \leq500GeV.
  \end{equation}

  \item We will focus on Higgs sector, so the stop sector is mainly used to tune the Higgs mass. We don't want a light stop to muss our analysis:
  \begin{equation}
      700GeV\leq M_{Q3}=M_{U3}\leq2TeV , \quad -3TeV\leq A_t\leq3TeV.
  \end{equation}

  \item $\mu$ correspond to Higgsino mass. One may let $\mu$ become large to avoid non-SM Higgs decaying to Higgsino. But we find a large $\mu$ always leading to a too heavy $M_A$ that almost touch the 2$m_t$ threshold(this can be understood by the pseudoscalar mass matrix(5)). So $\mu$ in our scenario is not too large :
  \begin{equation}
      100GeV\leq \mu \leq500GeV.
  \end{equation}

  \item All the other soft parameters will be set to 2TeV, and all the A-terms(except $A_t$, $A_\lambda$, $A_\kappa$) are equal to zero.

\end{itemize}

\subsection{Constraints}
Both the H1 and H2 could be the 125GeV SM-like Higgs.
So in this work, our discussion is performed in two scenarios:
\begin{itemize}
  \item H1 scenario: $H_1$ is the 125GeV SM-like Higgs.
  \item H2 scenario: $H_2$ is the 125GeV SM-like Higgs.
\end{itemize}

$\verb"NMSSMTools"$ have included lots of low-energy observables from B-physics, LEP, Tevatron, and LHC to constrain the parameter space.
Besides those traditional low-energy observables constraints, we describe some related constraints used in our work.
\begin{itemize}
  \item There is only one Higgs in 122GeV-128GeV(we have excluded the degenerate Higgs scenario in this work\cite{Gunion:2012gc}), and its Higgs signal strength should be consistent with Higgs data in 2$\sigma$ range.

  \item $\verb"HiggsBounds-4.1.2"$\cite{HiggsBounds} is used to constrain the Higgs sector.

  \item Non-SM Higgs searching through $H/A\rightarrow \tau\tau$ in CMS\cite{CMS_HIG_12_050}\cite{CMS_HIG_13_021} and ATLAS\cite{ATLAS-CONF-2014-049}, and through $H\rightarrow AA$ in CMS\cite{CMS_HIG_13_010} have been included in $\verb"NMSSMTools"$.
      In addition, the observed limit given in the non-SM Higgs search through $A\rightarrow Zh$ and $H\rightarrow hh$ is used to constrain the parameter space also.

  \item The thermal relic density of the lightest neutralino is required to be lower than the upper bound measured by Planck\cite{planck}, and the rescaled proton scattering cross section need to satisfy the direct detection bound from LUX\cite{lux}. The calculation is performed by $\verb"MicroOmega"$\cite{micro}.

  \item The lightest sfermion should be heavier than 500GeV.
\end{itemize}

Our scan is a multi-step scan. First, we scan the entire parameter space quickly, then we intensively scan the part that will leading to significant result.
Finally we sum all the partial results together. In next section we will show our scan results and analyse them in detail.

\section{Scan results}
In this section we will show our scan results and discuss why these plots are like that.
We will show the production crosssection of heavy scalars first, then analyse their branching ratio, finally the combination of crosssection and branching ratio will tell us how many 125GeV SM-like Higgs are produced from heavy scalars' decay.

\subsection{The Production Crosssection of Scalars}
The gluon fusion process $gg\rightarrow H$ and $gg\rightarrow A$ are calculated by HIGLU\cite{HIGLU} in NNLO.
While for vector boson fusion , top pair associated production, and associated production with vector boson, we just use the results in\cite{VH}.
And due to the $\tan\beta$ in our scenario is small, contribution from bottom quark fusion can be safely neglected.
For the production cross section of charged Higgs, we use the results in\cite{CHxs}.
Combined with the reduced couplings calculated by $\verb"NMSSMTools"$, we can get the production crosssection of all the 6 Higgs states.

\begin{figure}[t]
\mbox{\hspace*{-0.3cm}\includegraphics[clip=true,width=9.0cm]{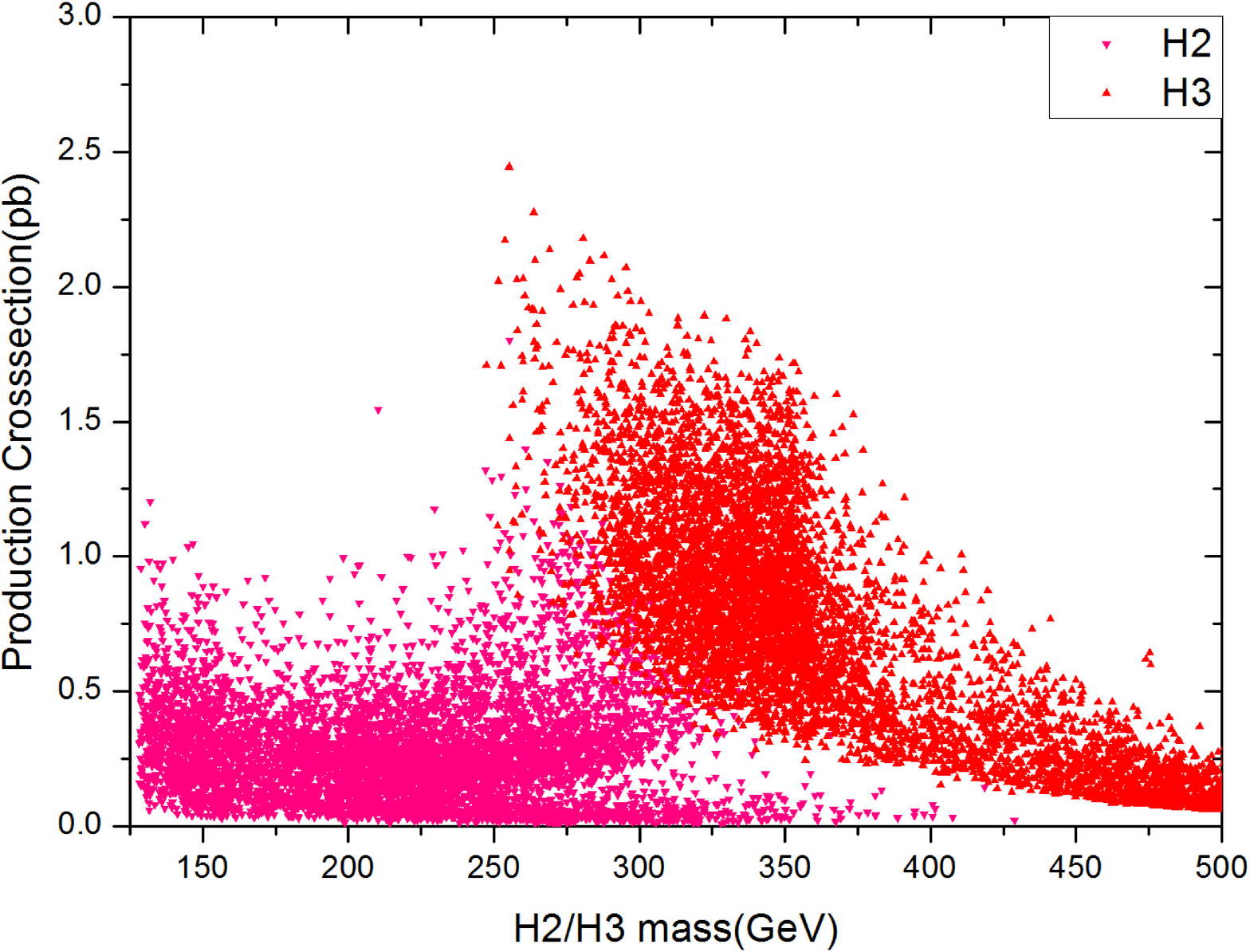}
\hspace*{-0.3cm}\includegraphics[clip=true,width=9.0cm]{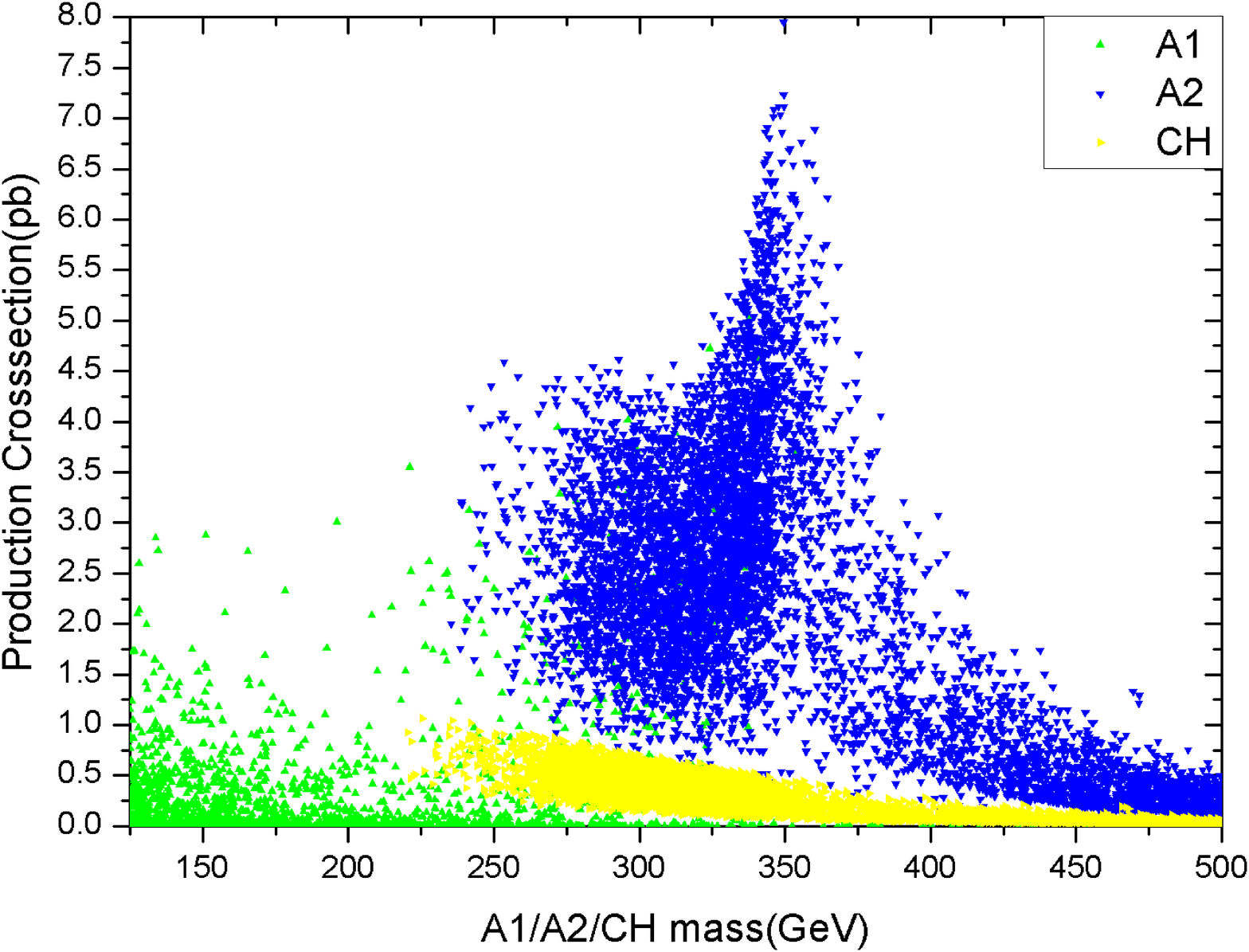}}
\caption{Left plot is the total production crosssection of $H_2$ and $H_3$ at 8TeV LHC, right plot is the total production crosssection of $A_1$, $A_2$, and $CH$ at 8TeV LHC. These results are obtained in H1 scenario($H_1$ is the 125GeV SM-like Higgs).}
\label{H1pro}
\end{figure}

\begin{figure}[t]
\mbox{\hspace*{-0.3cm}\includegraphics[clip=true,width=9.0cm]{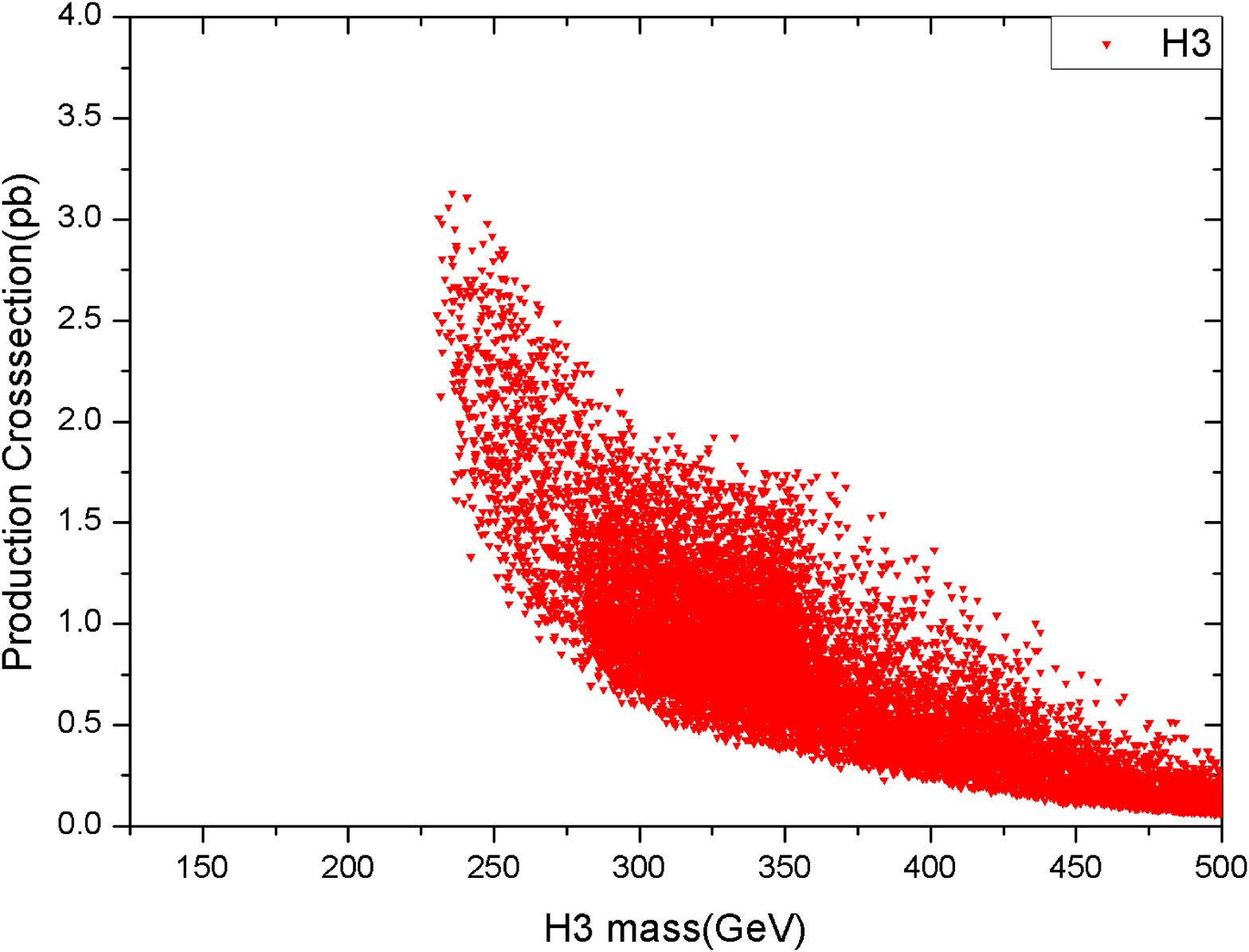}
\hspace*{-0.3cm}\includegraphics[clip=true,width=9.0cm]{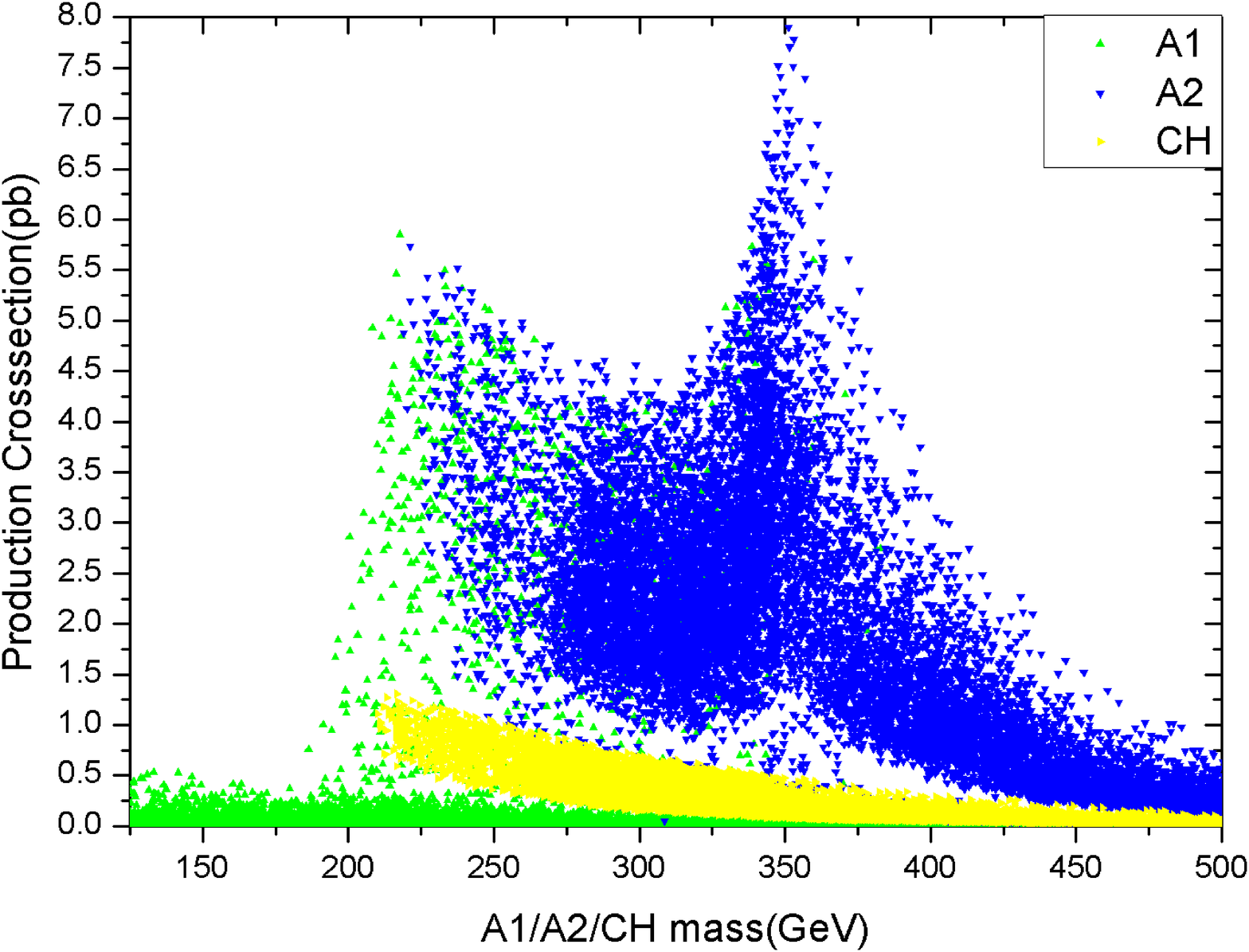}}
\caption{Left plot is the total production crosssection of $H_3$ at 8TeV LHC, right plot is the total production crosssection of $A_1$, $A_2$, and $CH$ at 8TeV LHC. These results are obtained in H2 scenario($H_2$ is the 125GeV SM-like Higgs).}
\label{H2pro}
\end{figure}

We want to explore how many 125GeV Higgs can decay from heavy scalars, so we will show the production crosssection of the scalars that are heavier than 125GeV only.
Fig.\ref{H1pro} is the total production crosssection of $H_2$, $H_3$, $A_1$, $A_2$, and $CH$ at 8TeV LHC in H1 scenario, and Fig.\ref{H2pro} is the total production crosssection of $H_3$, $A_1$, $A_2$, and $CH$ at 8TeV LHC in H2 scenario.
Compared to the total production crosssection of the 125GeV SM Higgs(about 20pb at 8TeV LHC\cite{VH}), the doublet dominant H3 and A2 still have a significant production crosssection.
The larger production crosssection of pseudoscalar is due to the fermion loop factor in Higgs-gluon-gluon effective coupling, and the peak around 2$m_t$ can be explained by the mixing effect of scalar and top quarkonium\cite{Drees:1989du}.
So, if the $H_3$ and $A_2$ branching ratio decaying to 125GeV Higgs are large, then such heavy Higgs decay effect may be important in Higgs signal analysis.
The singlet dominant $H_2$ and $A_1$ may also contribute to this effect if they mixing with the doublet adequately.
CH can contribute to this effect through $CH\rightarrow h W$(we will use $h$ to denote the 125GeV SM-like Higgs in below).
But the production crosssection of $CH$ is too small(because CH can't be produced through gluon-gluon fusion, its production will be suppressed by the bottom quark PDF and the associated produced top quark in the final state), and CH will mainly decay to top bottom(depend on $\tan\beta$)\cite{CHxs}.
So we don't expect a large contribution from $CH$.

\subsection{The Branching Ratios of Scalars}
Higgs SSB is the origin of SM fermion and vector boson mass.
So bottom pair, top pair, W boson pair and Z boson will be the main decay product of the neutral Higgs.
But the pseudoscalar can't couple to gauge boson in lowest order, so the pseudoscalar branching ratio to WW and ZZ can be ignored.
The Higgsino and Singlino in our scenario are not too heavy, so the branching ratio to $\tilde{\chi}^0$ and $\tilde{\chi}^\pm$ can not be ignored.
The sum of the branching ratio to $\tilde{\chi}^0$ and $\tilde{\chi}^\pm$ will be denoted as BRS:
\begin{eqnarray}
  BRS(H/A)&=& \sum_{i,j=1}^3 BR(\tilde{\chi}^0_i\tilde{\chi}^0_j) + BR(\tilde{\chi}^\pm_1) \\
  BRS(CH) &=& \sum_{i=1}^3 BR(\tilde{\chi}^0_i\tilde{\chi}^\pm_1)
\end{eqnarray}
In order to illustrate the heavy Higgs decay effect more obviously, we define a quantity "$Numh$".
$Numh$ is the average number of 125GeV SM-like Higgs produced from the heavy scalar decay.
For example, if $Numh(A_2)$ is 0.7, then 10000 $A_2$ will produce 7000 125GeV SM-like Higgs on average.
The cascade decay process like $A_2 \rightarrow H_2 Z \rightarrow hhZ$ also need to be included in $Numh$ calculation.
A detailed describe of $Numh$ calculation can be found in \cite{Higgs_Chain}\cite{Feeddown}.
We emphasize here that $Numh$ only include decay topology containing at least one 125GeV SM-like Higgs, but sometimes the heavy Higgs will tend to decay to non-SM Higgs.
In such situation $Numh$ will be small even the other branching ratios are small.

\begin{figure}[H]
\centering
\mbox{\hspace*{-0.3cm}\includegraphics[clip=true,width=9.0cm]{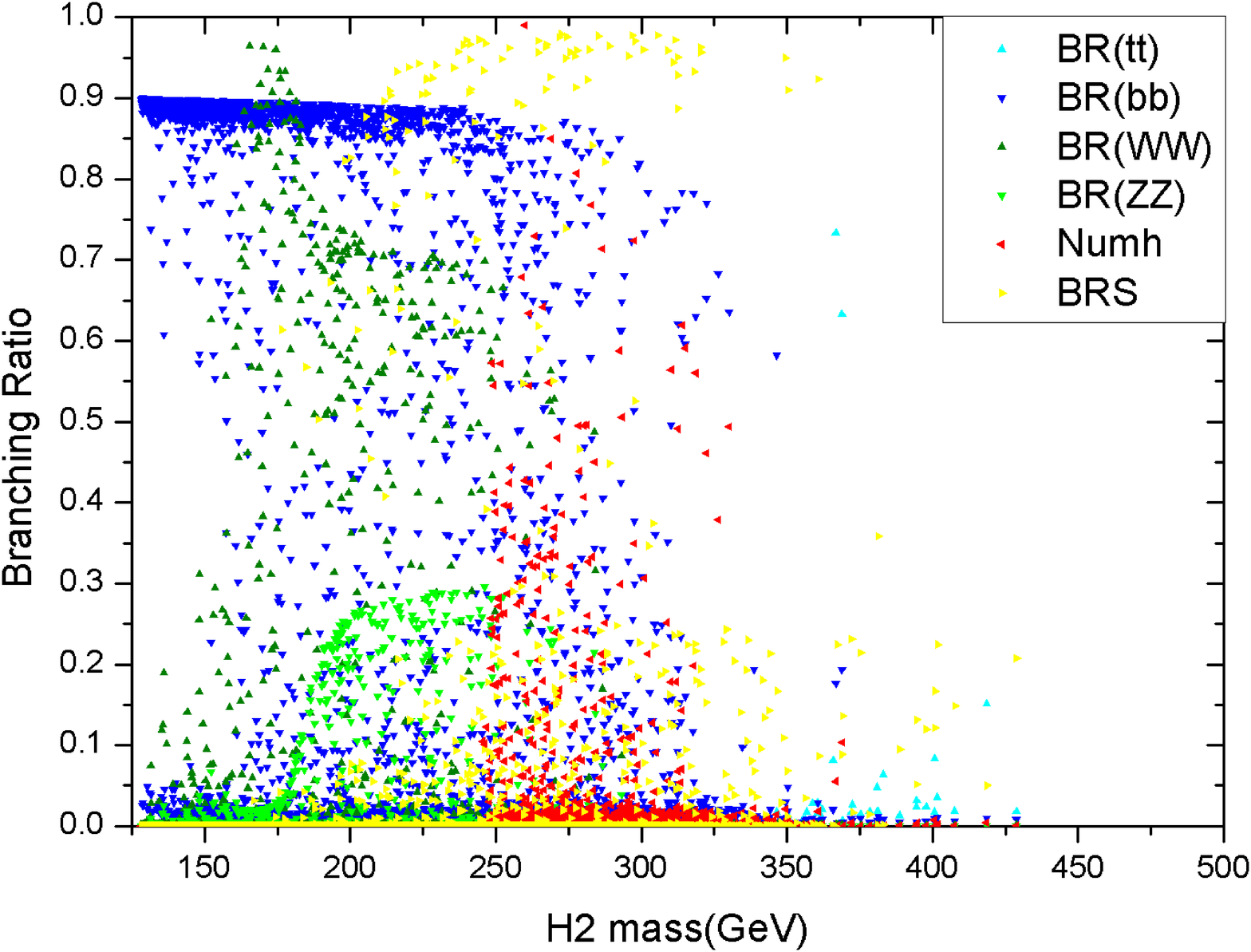}
\hspace*{-0.3cm}\includegraphics[clip=true,width=9.0cm]{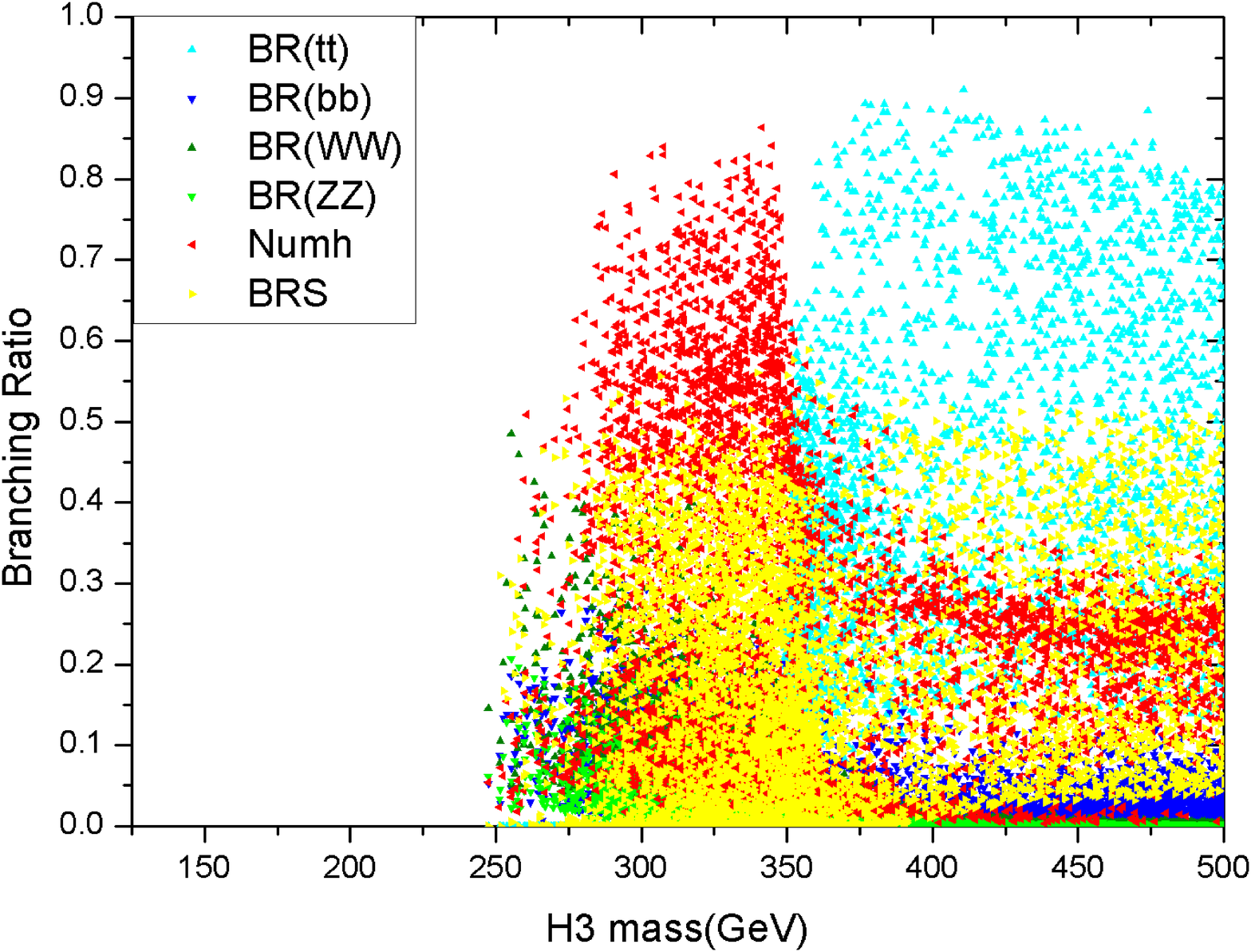}}
\mbox{\hspace*{-0.3cm}\includegraphics[clip=true,width=9.0cm]{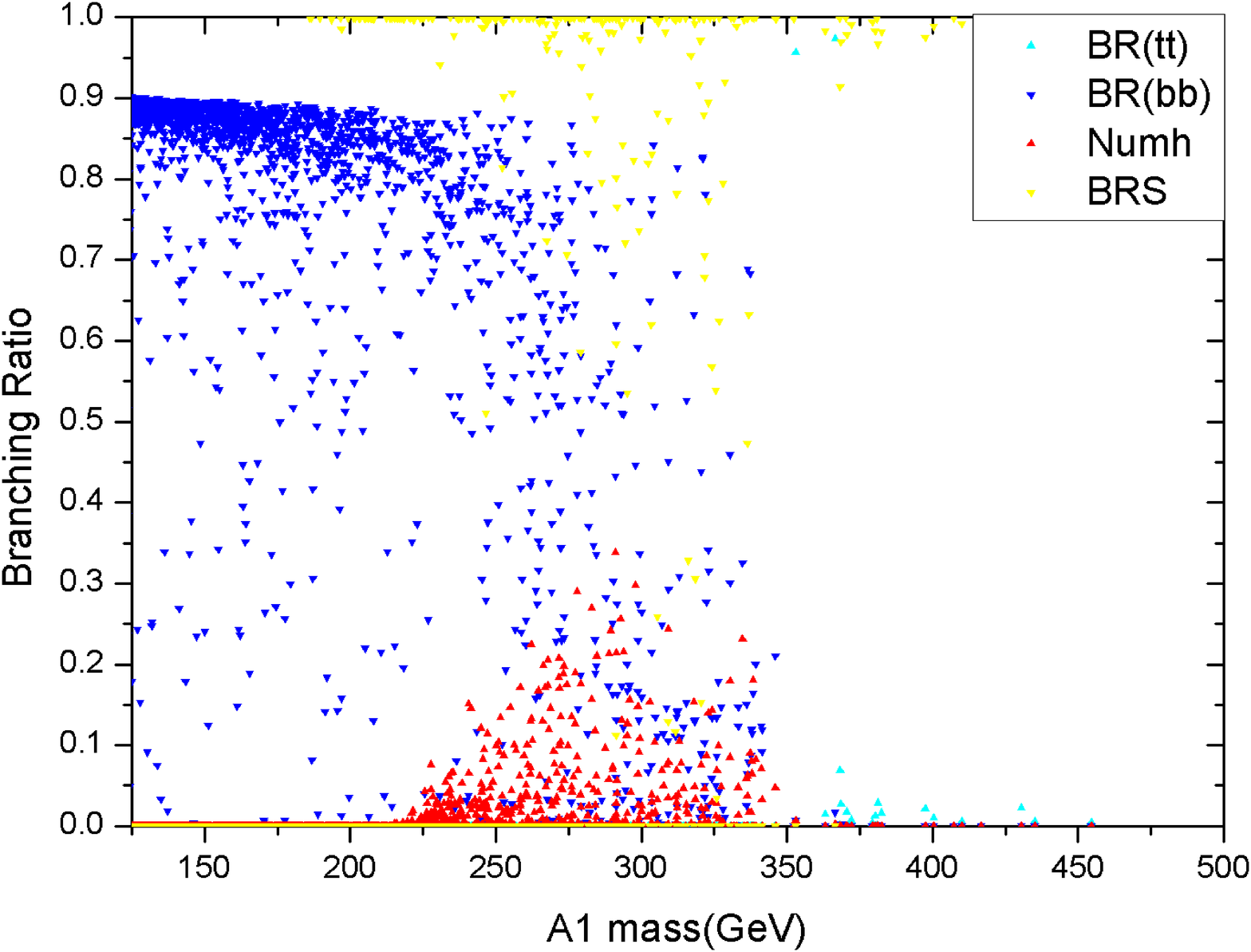}
\hspace*{-0.3cm}\includegraphics[clip=true,width=9.0cm]{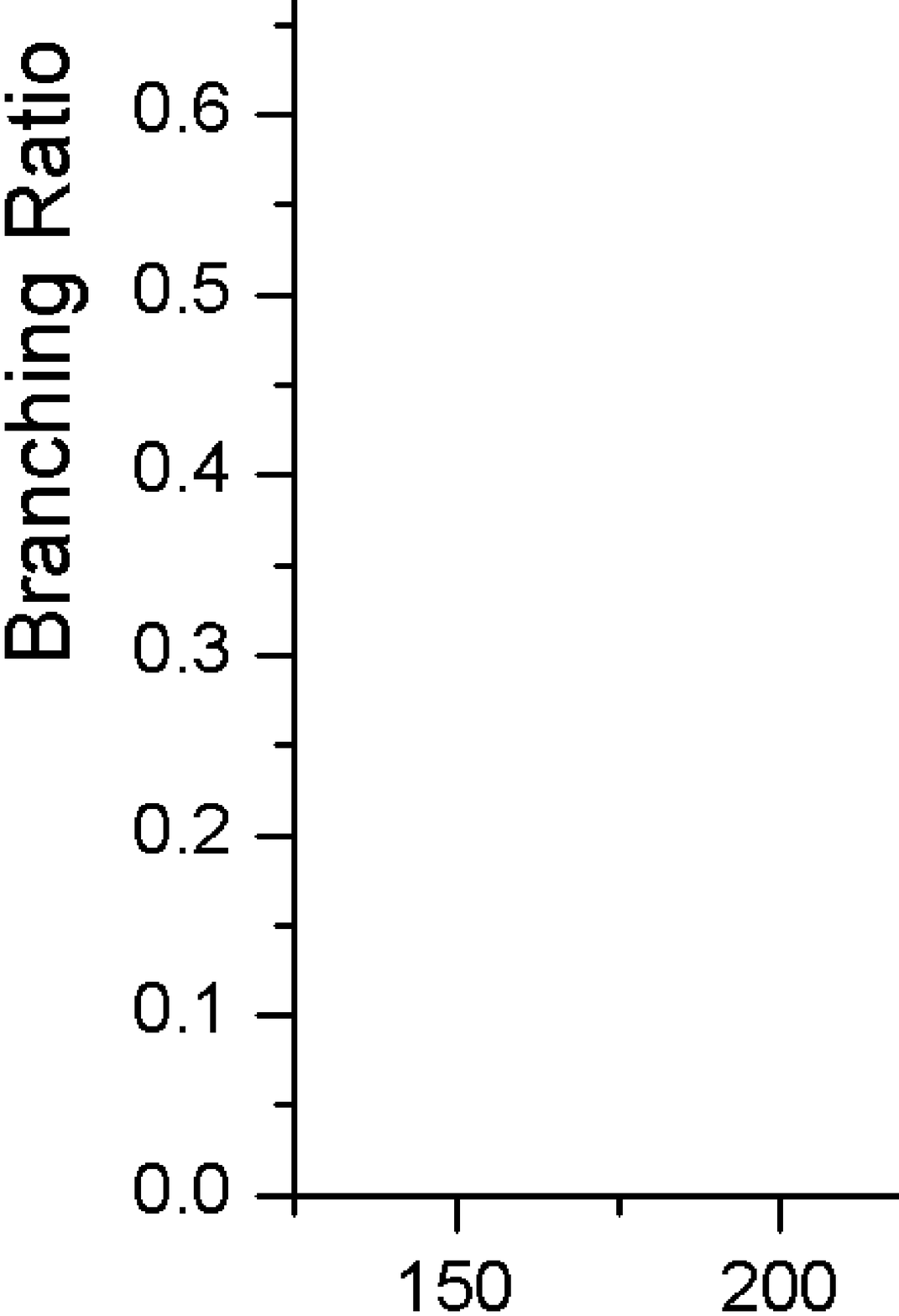}}
\mbox{\hspace*{-0.3cm}\includegraphics[clip=true,width=9.0cm]{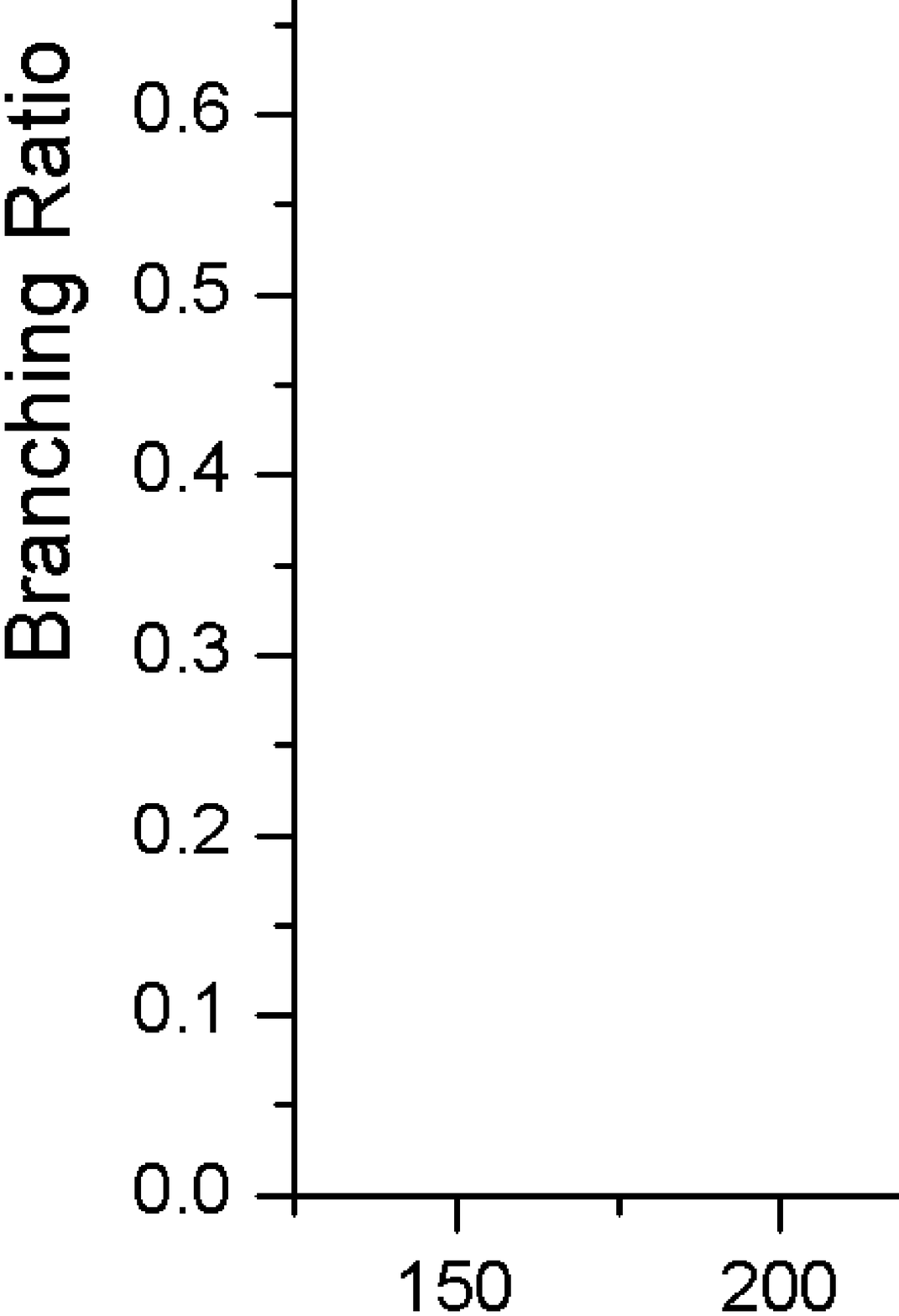}}
\caption{The main branching ratios of heavy Higgs in H1 scenario.}
\label{H1br}
\end{figure}

\begin{figure}[H]
\centering
\mbox{\hspace*{-0.3cm}\includegraphics[clip=true,width=9.0cm]{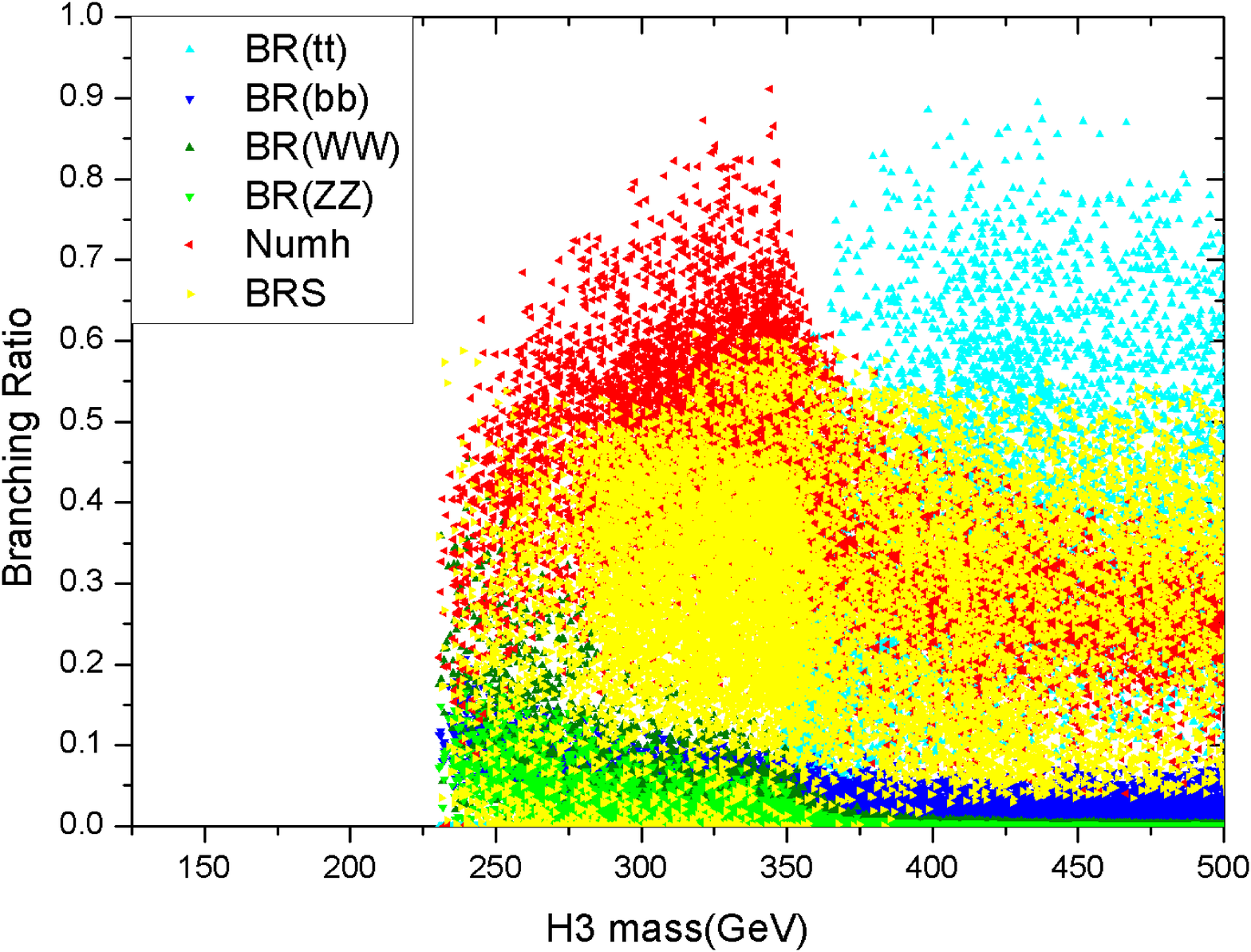}
\hspace*{-0.3cm}\includegraphics[clip=true,width=9.0cm]{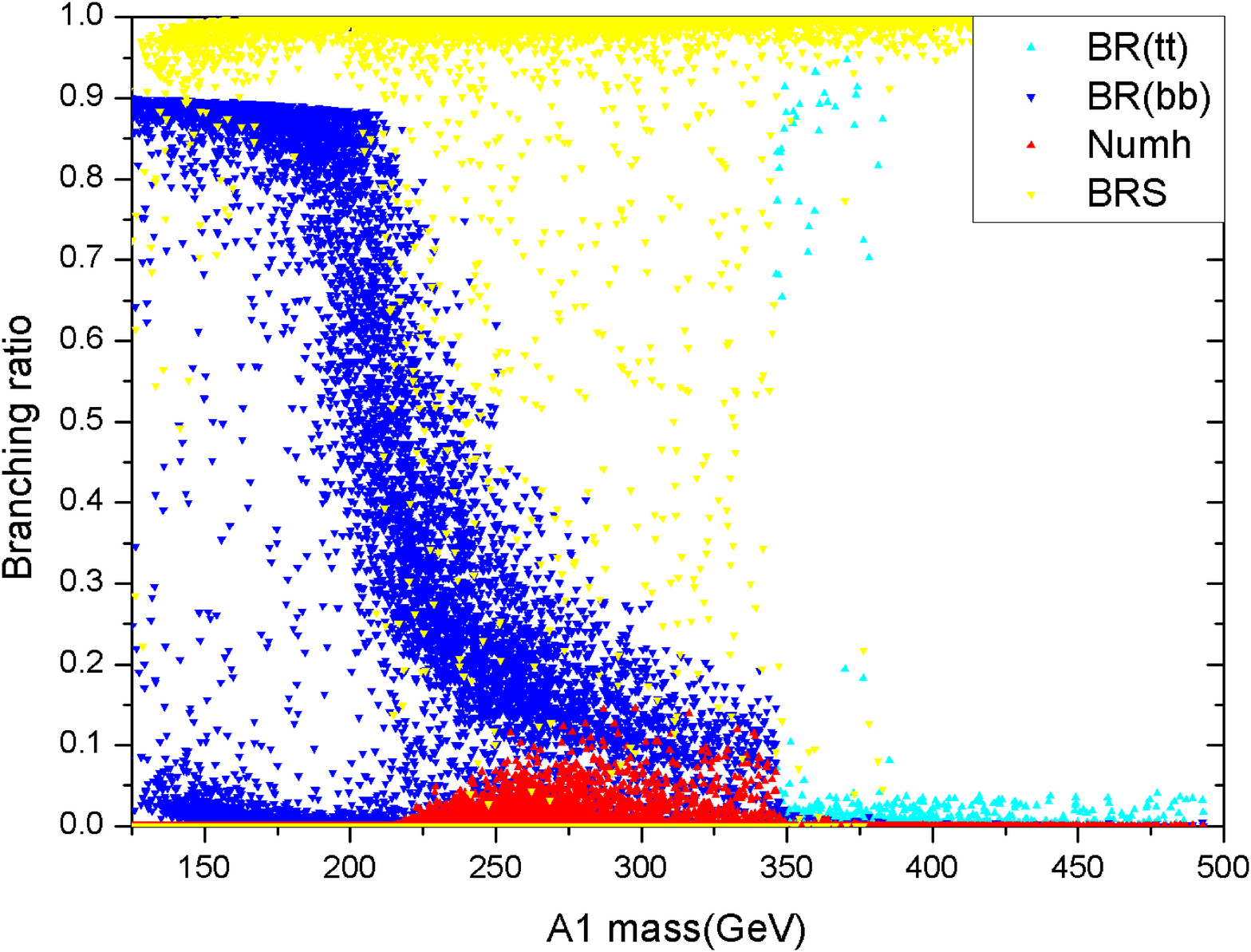}}
\mbox{\hspace*{-0.3cm}\includegraphics[clip=true,width=9.0cm]{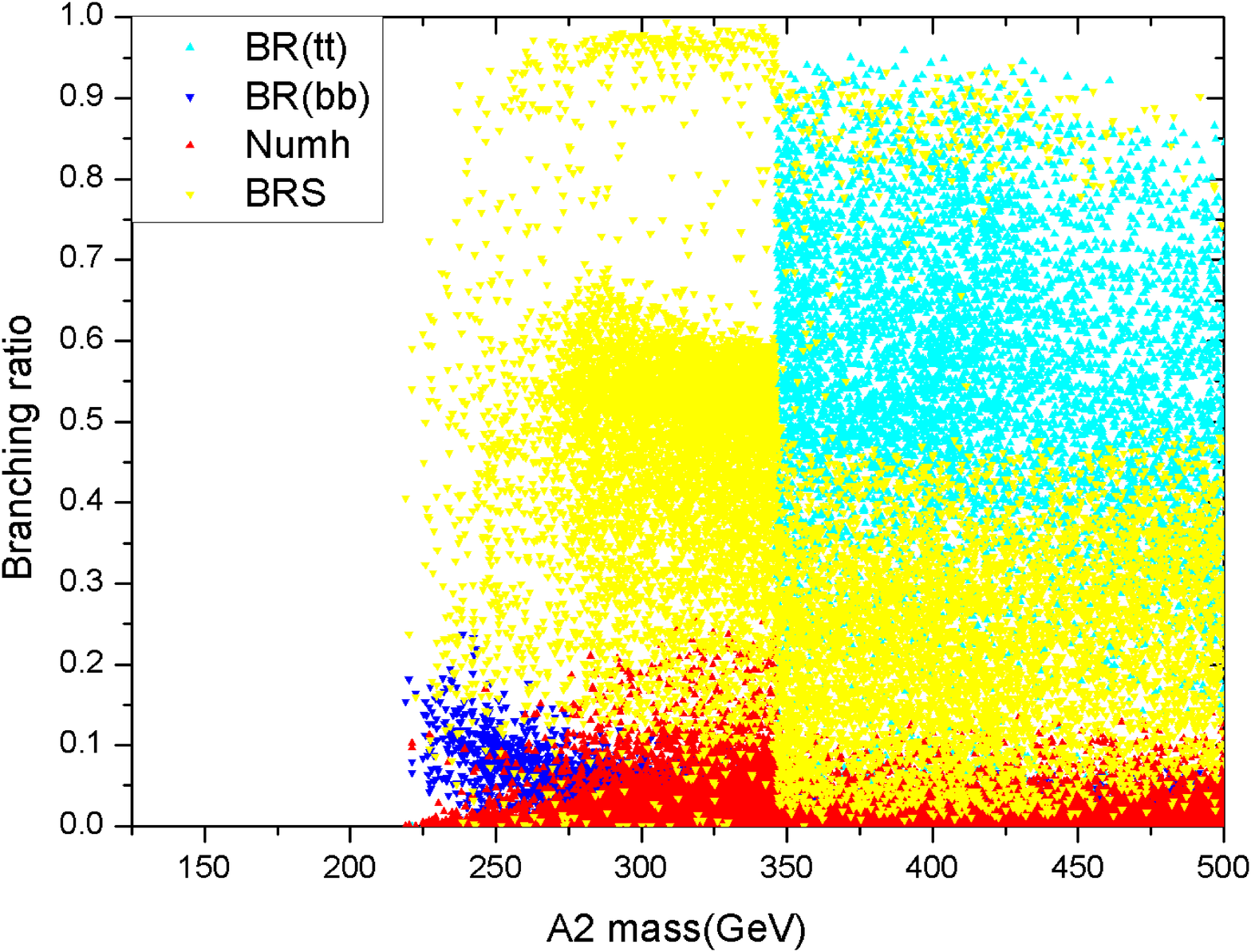}
\hspace*{-0.3cm}\includegraphics[clip=true,width=9.0cm]{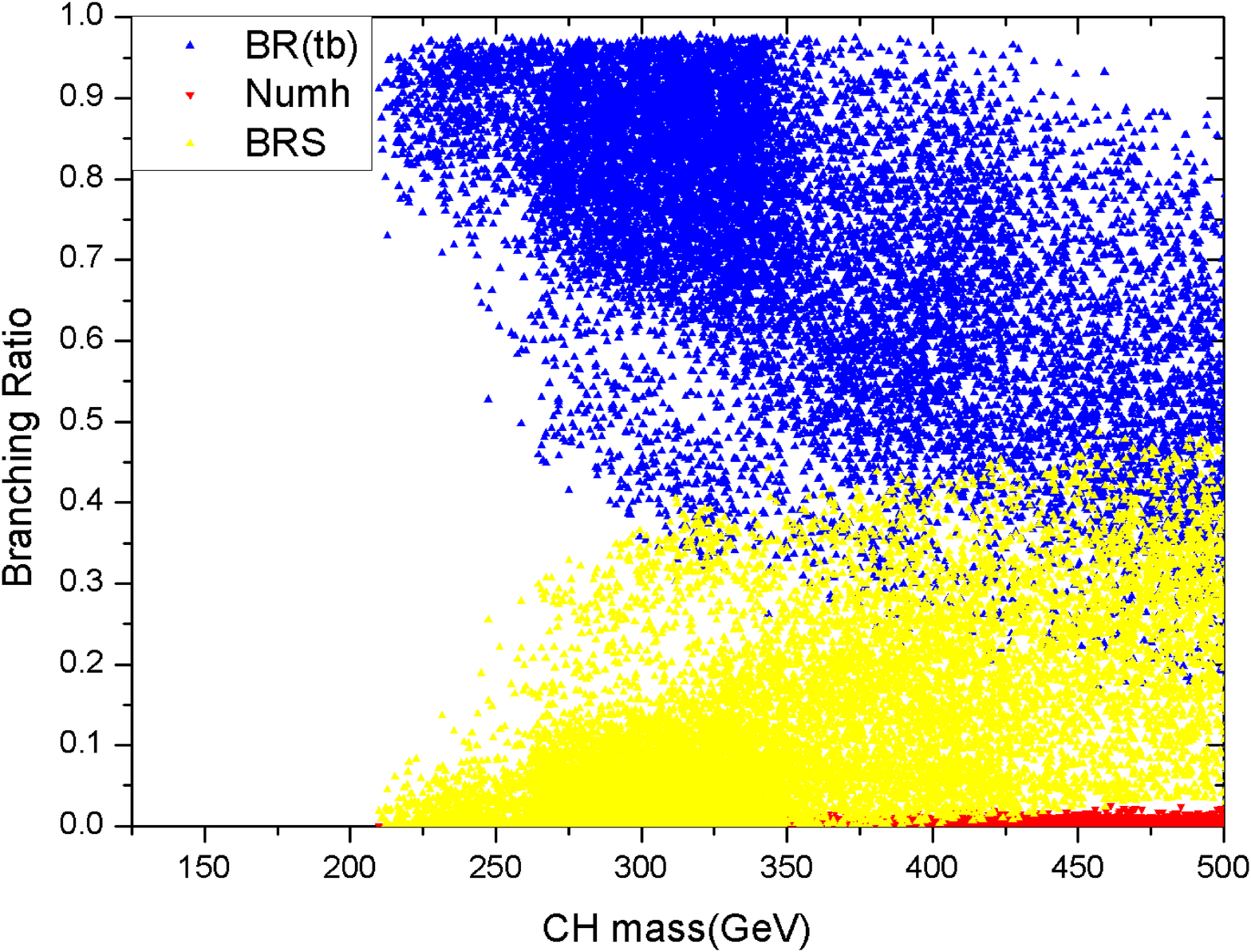}}
\caption{The main branching ratios of heavy Higgs in H2 scenario.}
\label{H2br}
\end{figure}

Fig.\ref{H1br} and fig.\ref{H2br} are the braching ratios of heavy Higgs in H1 and H2 scenario respectively.
A main property illustrated by these plots is that significant $Numh$ can only appear below the $2m_t$ threshold, as we mentioned before.
Light singlino and higgsino occupy a large part of decay product, especially in the H2 scenario.
$H_3$ always have a possibility to get a large $Numh$ before its mass reaching $2m_t$ threshold.
Due to a large production crosssection, $A_2$ can contribute largely to the heavy Higgs decay effect in H1 scenario.
While $A_2$ in H2 scenario tend to decay to non-SM Higgs, so the contribution from $A_2$ will be sub-dominant in H2 scenario.
$CH$'s branching ratio is mainly occupied by BR(tb) and BRS.
So combined with the production crosssection, contribution from CH will be very small.

We have production and decay of the NMSSM Higgs sector have been shown clearly, next we will show how much 125GeV Higgs can be produced from the heavy scalar's decay.

\subsection{Heavy Higgs Decay Effect in 125GeV Higgs production}
\begin{figure}[t]
\centering
\mbox{\hspace*{-0.3cm}\includegraphics[clip=true,width=9.0cm]{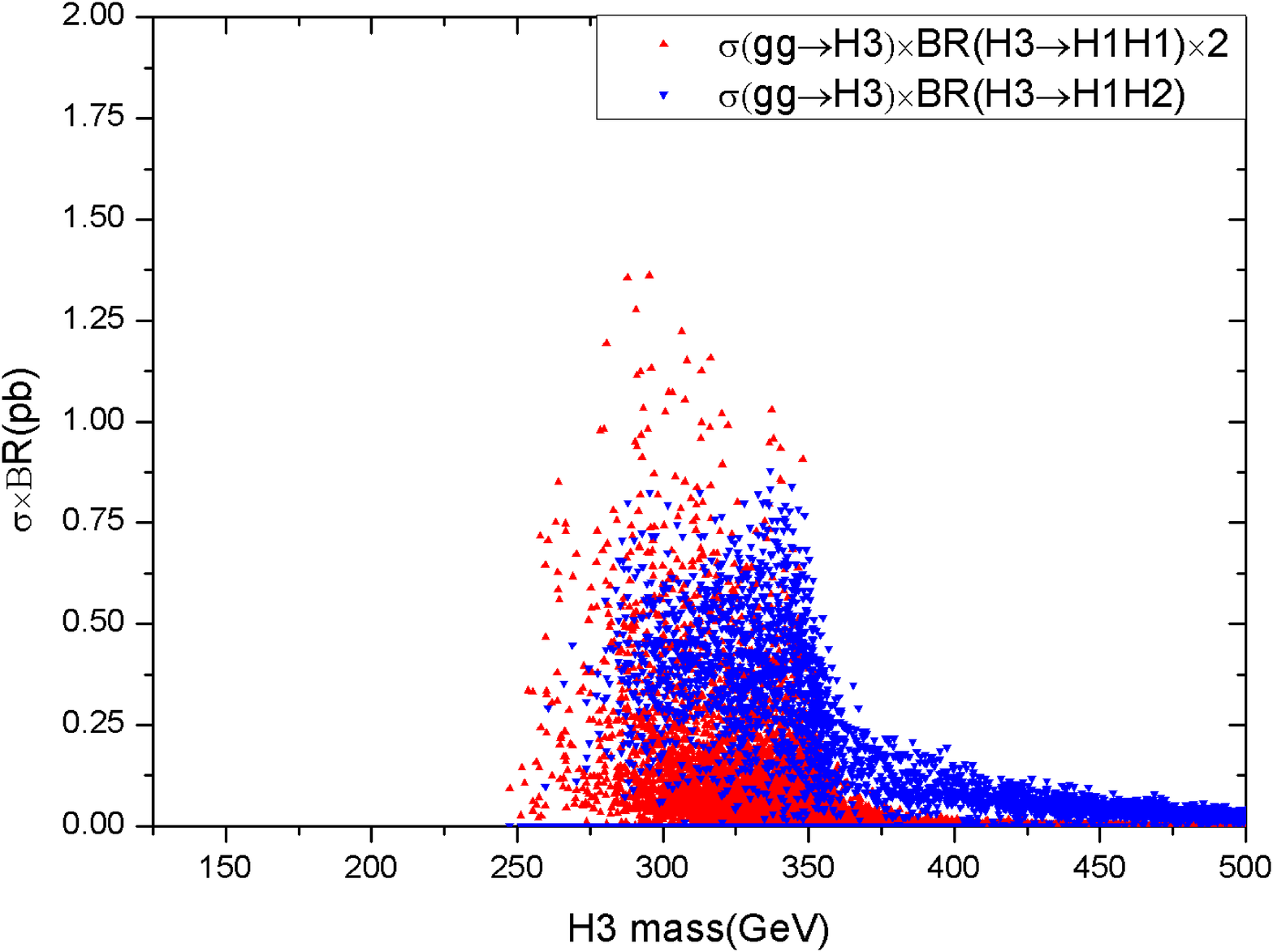}
\hspace*{-0.3cm}\includegraphics[clip=true,width=9.0cm]{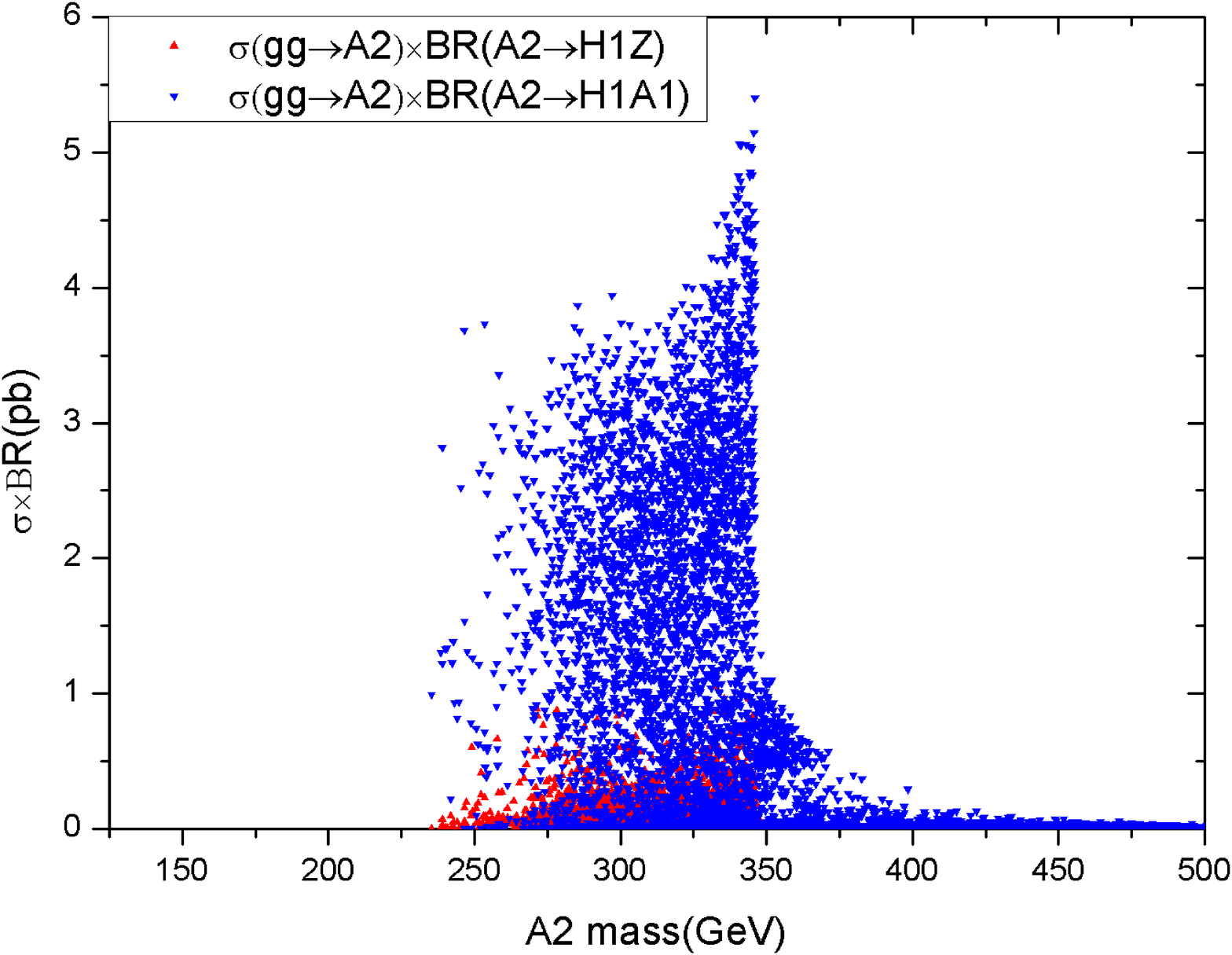}}
\caption{Channels that contribute significantly in H1 scenario. The gluon-fusion crosssection is obtained ar 8TeV LHC.}
\label{H1channel}
\end{figure}

\begin{figure}[t]
\centering
\mbox{\hspace*{-0.3cm}\includegraphics[clip=true,width=9.0cm]{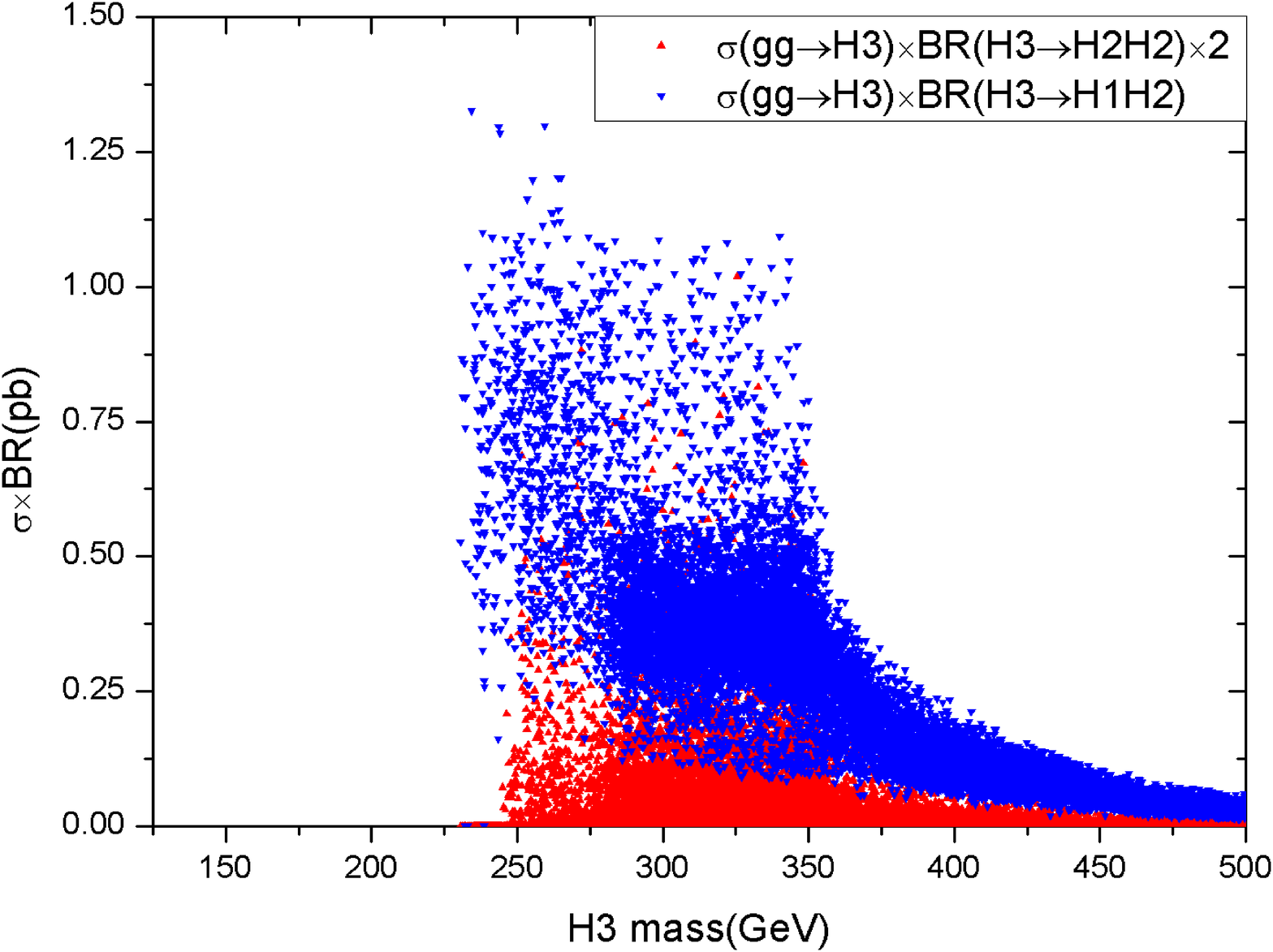}
\hspace*{-0.3cm}\includegraphics[clip=true,width=9.0cm]{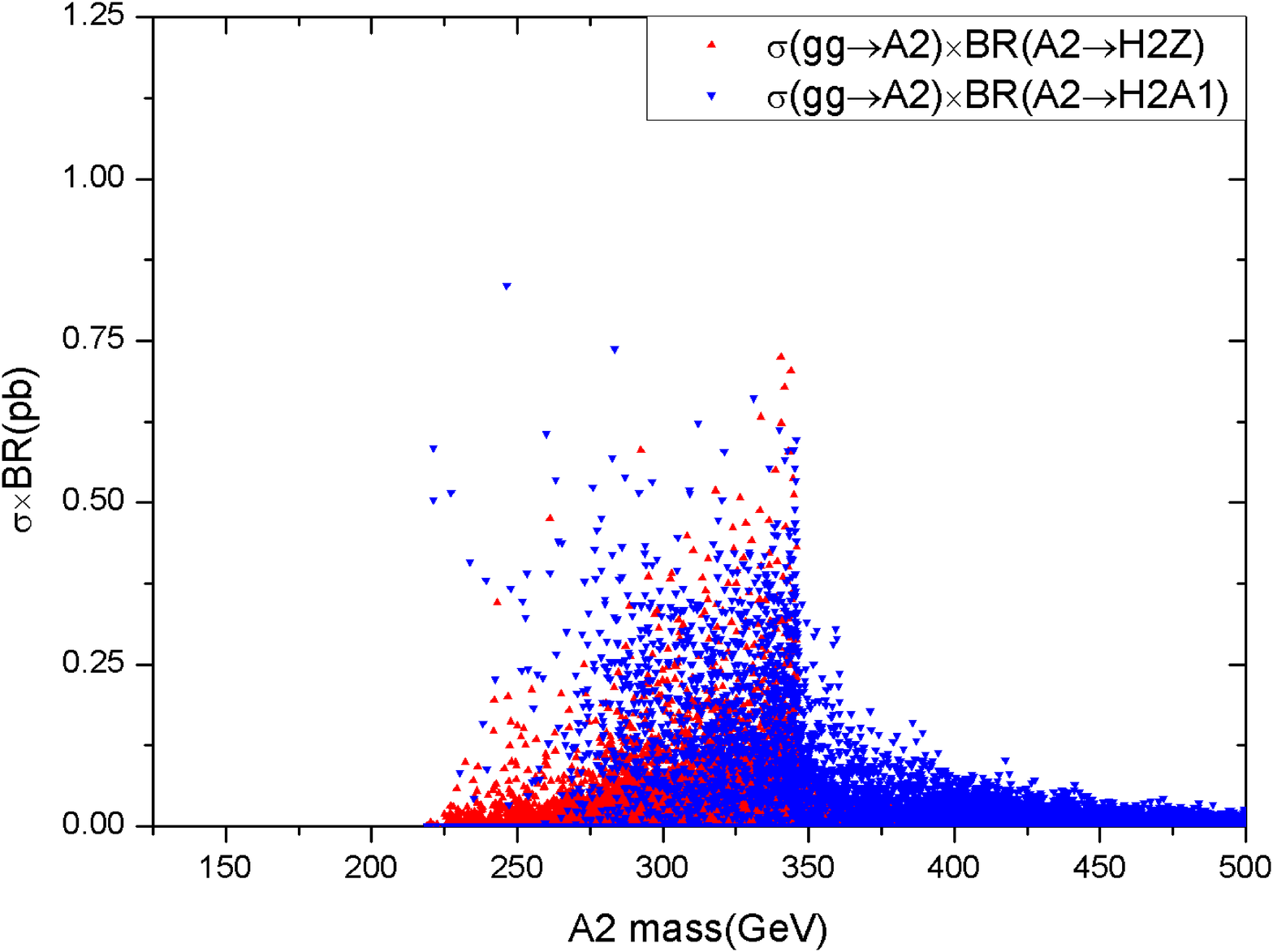}}
\caption{Channels that contribute significantly in H2 scenario. The gluon-fusion crosssection is obtained ar 8TeV LHC.}
\label{H2channel}
\end{figure}

Result can be simply obtained by multiplying production crosssection and $Numh$.
But before giving the final result, we will show the specific channel that have a significant contribution.
Besides the decay channel, the production process are separated also.
We find all the important contribution are from $H_3$ and $A_2$ in both scenarios.
Fig.\ref{H1channel} and Fig.\ref{H2channel} are the results.
Those processes could be search channel of NMSSM Higgs sector at LHC run II\cite{Kang}\cite{King:2014xwa}.
But we need to notice that such a significant crosssection is sensitive to the mixing between non-SM and SM Higgs, and the non-SM Higgs mass scale.
The mixing and mass scale allowed by current Higgs data will be excluded largely by future precise Higgs property measurement\cite{Dawson:2013bba}\cite{Wu:2015nba}\cite{Kakizaki:2015zva}.
So before we perform the NMSSM Higgs sector searching in future collider, these channels might be not promising anymore under updated constrain.

\begin{figure}[t]
\centering
\mbox{\hspace*{-0.3cm}\includegraphics[clip=true,width=9.0cm]{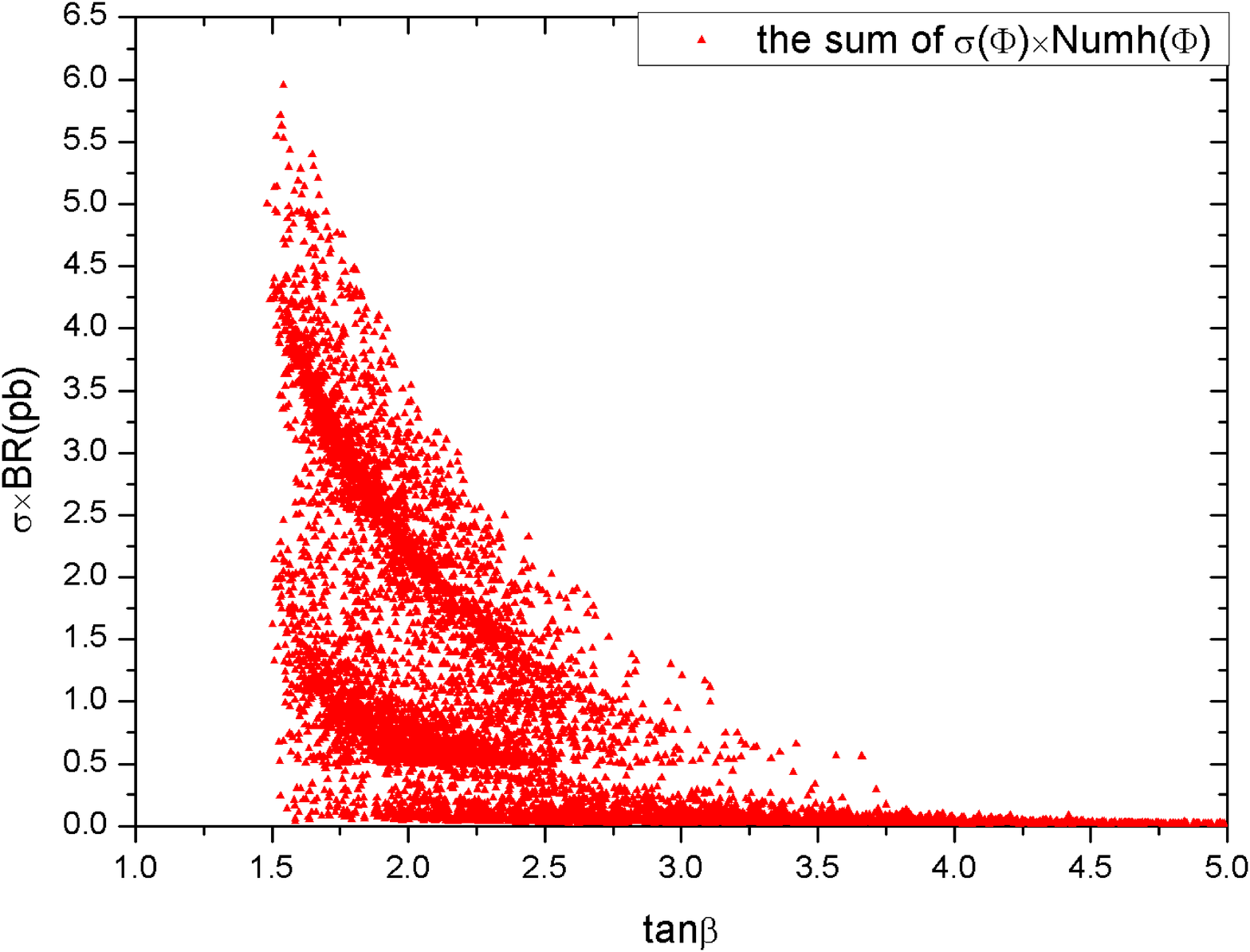}
\hspace*{-0.3cm}\includegraphics[clip=true,width=9.0cm]{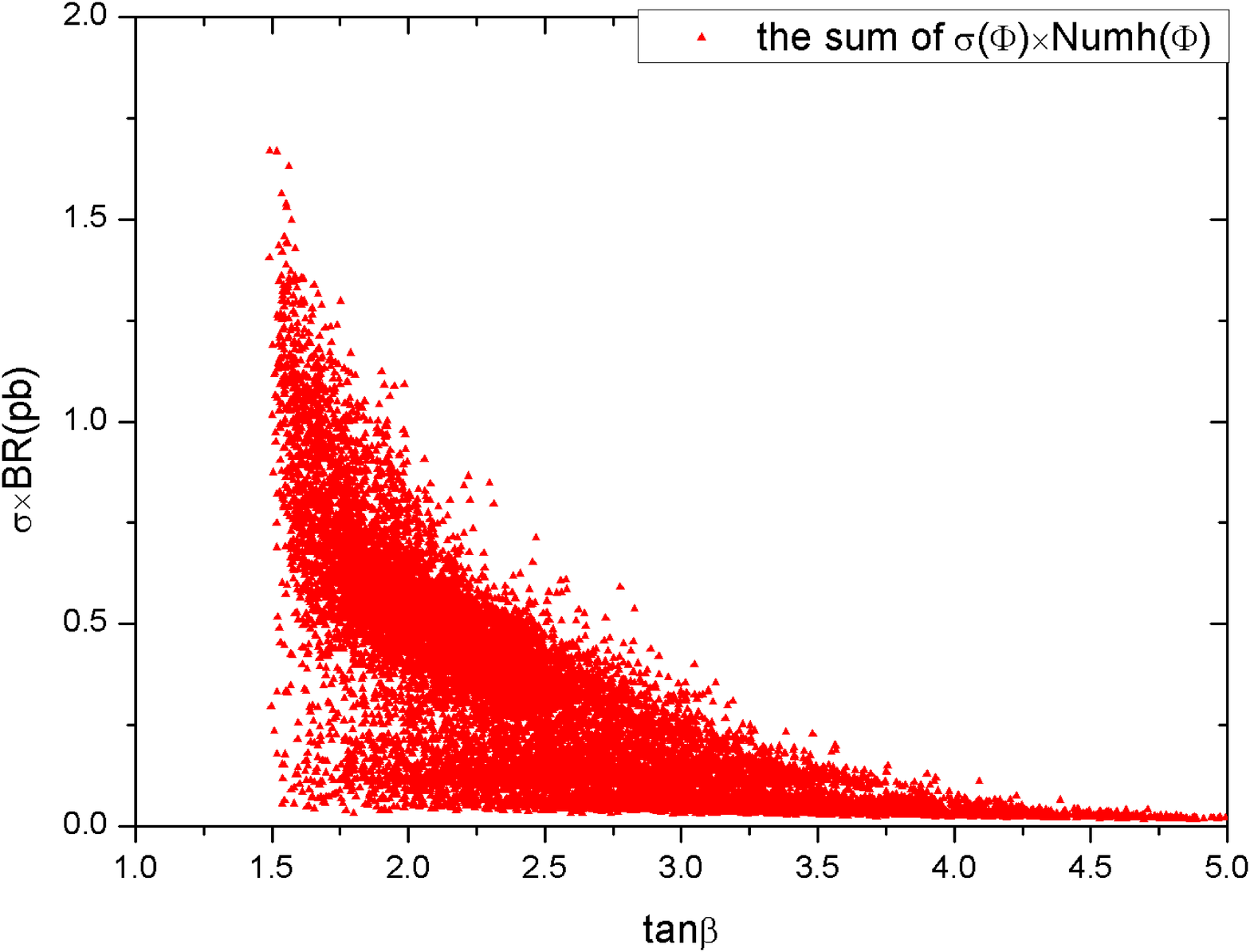}}
\caption{The sum of all production and decay channel of all the scalars that are heavier than 125GeV Higgs. Left plot is obtained in H1 scenario and right plot is obtained in H2 scenario.}
\label{num}
\end{figure}

Fig.\ref{num} is the sum of all production and decay channel for all the scalars that are heavier than 125GeV Higgs.
As we mentioned in introduction, this heavy Higgs decay effect is strongly suppressed by large $\tan\beta$.
This effect is much significant in H1 scenario that in H2 scenario.
In next section, we will analyse how this effect change Higgs data fitting.

\section{The Impact on Higgs Data Fitting}
In the final section, we will show how the heavy Higgs decay effect affect the Higgs data fitting.
Considering the uncertainty in the gluon-gluon fusion process calculation is larger than 15\% and the production crosssection of SM Higgs in 8TeV LHC is about 20pb\cite{uncertainty}, we will focus on H1 scenario only.
Let's briefly review the Higgs data fitting first.

The data used to fit is the Higgs signal strength $\mu$ which is the ratio of observed signal event number to the SM expected number.
While the NP models may lead to a $\mu$ that not equal to 1, so the observed $\mu$ can be used to constrain the NP models.
$\mu$ can be calculated by production crosssection, branching ratio, and signal efficiency:
\begin{equation}\label{mu}
  \mu(X,Y)\equiv \frac{\sigma(X)BR(h\rightarrow Y)\epsilon}{\sigma_{SM}(X)BR_{SM}(h\rightarrow Y)\epsilon_{SM}}.
\end{equation}
Where X is production process, Y is the final state, $\epsilon$ and $\epsilon_{SM}$ are the signal efficiency in NP model and SM respectively.
If the non-SM Higgs production process can be ignored, then $\epsilon$ and $\epsilon_{SM}$ can be ignored too, because they will be just the same\cite{Bernon:2015oza}.

In our H1 scenario, the heavy Higgs decay effect can not be ignored(larger than 5pb), so the calculation of $\mu$ will be different.
In such situation, signal efficiency can not be ignored again, we need to know how the non-SM Higgs production process feed each of the signal region measured by the experiment.
Dividing the numerator and denominator simultaneously, $\mu$ will be:
\begin{equation}
  \mu(X,Y)\equiv \frac{\sigma(X)BR(h\rightarrow Y)+\sigma(HD)BR(h\rightarrow Y)\frac{\epsilon}{\epsilon_{SM}}}{\sigma_{SM}(X)BR_{SM}(h\rightarrow Y)}.
\end{equation}
"$HD$" denote heavy Higgs decay, and we will use HD for short in the remaining part.
It's obviously that the core problem in here is to decide the ratio of efficiency $\frac{\epsilon}{\epsilon_{SM}}$, so a detailed collider simulation is needed.
But we do not prepare to perform a collider simulation in this work.
We will just suppose several different $\frac{\epsilon}{\epsilon_{SM}}$ and show how the Higgs data fitting change.
The decision of $\frac{\epsilon}{\epsilon_{SM}}$ by detailed simulation will be our future work.
By the way, the report released recently\cite{Aad:2015lha} show a discrepancy between the experimental observation and theoretical prediction in Higgs process,
so a precise simulation seems not easy.
On the other hand, our fitting process will not be performed to the exclusive signal region like\cite{Higgs_fit}.
We will use the combined result of ZZ and $\gamma\gamma$ channel given by the Fig.2 of \cite{HiggsUpdatedMeasurementATLAS} to perform the fitting.
Then $\chi^2$ can be calculated simply by:
\begin{equation}
  \chi^{2} = \frac{(\mu_{\gamma\gamma} - \hat{\mu}_{\gamma\gamma})^2}{\Delta^2_{\gamma\gamma}} + \frac{(\mu_{ZZ} - \hat{\mu}_{ZZ})^2}{\Delta^2_{ZZ}}.
\end{equation}
Where the hatted number and $\Delta$ are the central value and error bar given by Fig.2 of \cite{HiggsUpdatedMeasurementATLAS}.

Next we will show the $\chi^2$ distribution in $\tan\beta-M_A$ plane in three scenarios: $HD$ is ignored; $\frac{\epsilon}{\epsilon_{SM}}$ is 0.4; $\frac{\epsilon}{\epsilon_{SM}}$ is 0.8.
But, could the ratio of efficiency $\frac{\epsilon}{\epsilon_{SM}}$ be as large as 0.8 or 0.4?
There are two reason for it:
\begin{itemize}
  \item The mass splitting between $M_{A_2}$ and $M_{h}+M_{A_1}$, or $M_{H_3}$ and $M_{h}+M_{H_1}/M_{H_2}$ are not too large, so the differential distribution of the 125GeV Higgs produced from $HD$ will not be too different from the 125GeV Higgs produced from SM processes.
  \item The main decay product of $A_1$, $H_1$, and $H_2$ are bottom pair or neutralino, but the description of the inclusive signal of ZZ and $\gamma\gamma$ channel(section 5.1 of \cite{Aad:2014eva} and section V of \cite{Aad:2014eha}) do not mention requirement on b-jet or missing energy.
\end{itemize}
Due to these two reasons, we think $HD$ will feed the Higgs searching signal effectively.
(We need to emphasize again that our work just give a rational assumption, the precise value of $\frac{\epsilon}{\epsilon_{SM}}$ can only be obtained by collider simulation.)

\begin{figure}[t]
\centering
\includegraphics[clip=true,width=17.0cm]{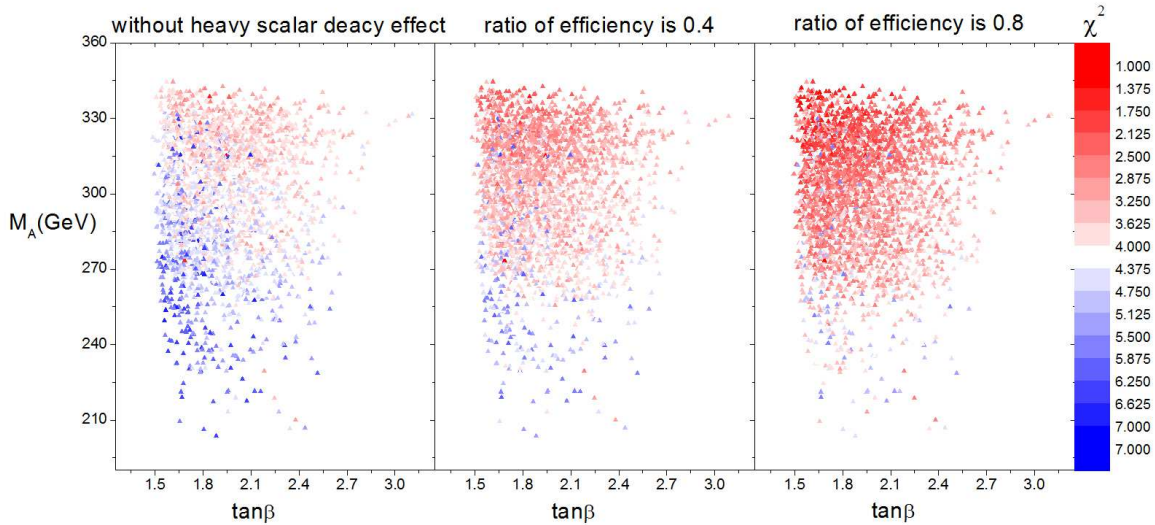}
\caption{The $\tan\beta-M_A$ plane }
\label{fit}
\end{figure}

Fig.\ref{fit} is the comparison of these three scenarios.
In order the illustrate the $HD$ effect more clearly, we pick the points with $HD$ effect larger than 1.0pb.
The color pattern show a clear changing in Higgs signal strength fitting.
Due to $BR(h\rightarrow ZZ)$ and $BR(h\rightarrow \gamma\gamma)$ suppression (because $hbb$ will be enhanced by mixing with heavy doublet), and the mixing with singlet, the signal strength of $h$ in NMSSM always tend to be smaller than the SM expected value.
So the $HD$'s feeding always lead to a smaller $\chi^2$, or say higher confidence level.

\section{Conclusion}
Higgs signal strength fitting always ignore the signal efficiency.
But this method is right only if the non-SM Higgs production is very small.
In this work, we find $A\rightarrow Zh$, $H\rightarrow hh$, $A_2\rightarrow A_1h$, and $H_3\rightarrow H_1H_2$ can contribute largely to the 125GeV SM-like Higgs production.
The corresponding total crosssection can be large than 6.0pb.
Then we discuss how such a non-SM Higgs production process will change the Higgs data fitting.
By a reasonable assumption to the signal efficiency, the fitting result change largely by the feeding of this non-SM Higgs production process.
The precise estimation of this effect need detailed collider simulation which beyond the scope of this work.

\vspace{0.2cm}

{\bf{Acknowledgement:}}
Thanks Jin Min Yang, Lei Wu, and Chengcheng Han for their great help and fruitful discussion.
This work was supported by the National Natural Science Foundation of China (NNSFC) under grants Nos. 11275245.


\begin{thebibliography}{99}
\bibitem{HiggsUpdatedMeasurementCMS}
  CMS Collaboration [CMS Collaboration],
  ``Precise determination of the mass of the Higgs boson
and studies of the compatibility of its couplings with the standard model,''
  CMS-PAS-HIG-14-009.
\end{thebibliography}
\end{document}